\documentclass[12pt]{article}
\usepackage{amsmath}
\usepackage{amssymb}
\usepackage{graphicx}
\usepackage{float}
\usepackage{indentfirst}
\usepackage{mathrsfs}

\usepackage{setspace}

\usepackage[labelfont=bf,labelsep=period]{caption}

\usepackage[numbers,sort&compress]{natbib}

\newtheorem{theorem}{Theorem}
\newtheorem{proposition}{Proposition}

\title{Reformulating the Quantum Uncertainty Relation}

\author
{Jun-Li Li$^{1}$ and Cong-Feng Qiao$^{1,2}$\footnote{corresponding author;
qiaocf@ucas.ac.cn}\\ [0.2cm]
\normalsize{$^1$Department of Physics, University of the Chinese Academy of Sciences,}\\
\normalsize{YuQuan Road 19A, Beijing 100049, China}\\[2pt]
\normalsize{$^2$CAS Center for Excellence in Particle Physics, Beijing 100049, China}
}

\date{}

\begin{document}
\baselineskip24pt \maketitle

\begin{abstract}

Uncertainty principle is one of the cornerstones of quantum theory. In the
literature, there are two types of uncertainty relations, the operator
form concerning the variances of physical observables and the entropy form
related to entropic quantities. Both these forms are inequalities
involving pairwise observables, and are found to be nontrivial to
incorporate multiple observables. In this work we introduce a new form of
uncertainty relation which may give out complete trade-off relations for
variances of observables in pure and mixed quantum systems. Unlike the
prevailing uncertainty relations, which are either quantum state dependent
or not directly measurable, our bounds for variances of observables are
quantum state independent and immune from the ``triviality'' problem of
having zero expectation values. Furthermore, the new uncertainty relation
may provide a geometric explanation for the reason why there are
limitations on the simultaneous determination of different observables in
$N$-dimensional Hilbert space.

\end{abstract}

\section{Introduction}

The uncertainty principle is one of the most remarkable characteristics of
quantum theory, which the classical theory does not abide by. The first
formulation of the uncertainty principle was achieved by Heisenberg
\cite{Heisenberg-o}, that is the renowned inequality $\Delta x\Delta p\geq
\hbar/2$, which comes from the concept of indeterminacy of simultaneously
measuring the canonically conjugate quantities position and momentum of a
single particle. Later, Robertson generalized this uncertainty relation to
two arbitrary observables $A$ and $B$ as \cite{Robertson}
\begin{eqnarray}
\Delta A \Delta B \geq |\langle C \rangle| \; ,
\label{Robertson}
\end{eqnarray}
where the standard deviation, i.e. the square root of the variance, is
defined to be $\Delta X \equiv \sqrt{\langle X^2\rangle -\langle
X\rangle^2}$ for observables $X$ and the commutator $2iC = [A,B] \equiv
AB-BA$. The relation (\ref{Robertson}) reflects that the standard deviations
of $A$ and $B$ are bounded by the expectation value of their commutator in
a given quantum state of the system. However, this expectation value can
be zero, even for observables that are incompatible, and makes the
inequality trivial. To address this problem the uncertainty relation was
expressed in terms of Shannon entropies \cite{Deutsch-entropy,entropic-un},
where an improved version takes the following form \cite{Maassen-Uffink}
\begin{eqnarray}
H(A)+H(B) \geq -2\log c_{ab}\; .
\end{eqnarray}
Here $H(A)$ is the Shannon entropy of the probability distribution of the
eigenbasis $\{|a_m\rangle\}$ in the measuring system, and similarly is the
$H(B)$. The bound $c_{ab} \equiv \mathrm{Max}_{m,n}|\langle a_m|b_n\rangle|$
is the eigenbases' maximum overlap of operators $A$ and $B$, and therefore
is independent of the state of system.

The progress in the study of uncertainty relation has profound significance
on the formalism of quantum mechanics(QM) and far reaching consequences in
quantum information sciences, e.g., providing the quantum separability
criteria \cite{uncertainty-sep}, determining the quantum nonlocality
\cite{entropic-nonlocality,ascertaining-correlation} (see for example
Ref.~\cite{Entropy-survey} for a recent review). Therefore, the uncertainty
principle has been the focus of modern physics for decades.

For variance-based uncertainty relations, improvements designated for mixed
states had been proposed with strengthened but state dependent lower bounds
\cite{Mixed-improve,geo-state}. Despite the progress on getting stronger
uncertainty relations \cite{Schrodinger-uncertainty, Pati-uncertainty}, the
problem that the lower bounds depend on the state of the system remains
\cite{Bannur}. On the other hand, by proposing new measures of uncertainties
similar as that of entropy, state independent lower bounds could be obtained
\cite{geo-ent}. There was also the combination approach involving both the
entropic measures and variances in a single uncertainty relation, where only
a nearly optimal lower bound could be derived \cite{Variance-entropic}.
Hence, obtaining the state independent optimal trade-off uncertainty
relation for variances of physical observables is still an urgent and open
question.

In this work, we present a new type of uncertainty relation for multiple
physical observables, which is applicable in cases of both pure and mixed
quantum states. Our strategy to obtain the uncertainty relation includes
three steps: first decompose the quantum state of system and physical
observables in Bloch space; then express the variances of observables as
functions of relative angles between Bloch vectors; last, apply triangle
inequalities to these angles to get constraint functions for the variances
of observables, which may remarkably give out the state independent optimal
trade-off uncertainty relations.

\section{Reformulating the uncertainty relation}

\subsection{ Variances in form of Bloch vectors}

In quantum theory, the systems are generally described by density matrices,
which are Hermitian, and physical observables are represented by operator
matrices, which are also Hermitian. The uncertainty of an observable $A$ for
the physical system represented by matrix $\rho$ is measured by the variance
\begin{eqnarray}
\Delta A^2 = \mathrm{Tr}[A^2\rho] -
\mathrm{Tr}[A\rho]^2 \; . \label{Eq-variance-def}
\end{eqnarray}
Here Tr is the trace of a matrix. Note that the variance $\Delta A^2$ are
invariant under a substraction of constant diagonal matrix from $A$, i.e.,
$\Delta (A-\alpha I)^2 = \Delta A^2$ with $\alpha$ being any real number and
$I$ is the identity matrix. Therefore, without loss of generality we are legitimate to consider observables of traceless Hermitian operators.

The $N\times N$ unitary matrices with determinant 1 form the special unitary
group of degree $N$, denoted by SU($N$). There are $N^2-1$ traceless
Hermitian matrices $\lambda_j$ of dimension $N\times N$, which constitute
the generators of SU($N$) group
\begin{eqnarray}
[\lambda_{j},\lambda_{k}] = 2i\sum_{l=1}^{N^2-1} f_{jkl}\lambda_{l} \; , \;
\{\lambda_{j},\lambda_{k}\} = \frac{4}{N}\delta_{jk}I +
2\sum_{l=1}^{N^2-1} d_{jkl} \lambda_{l} \; ,
\end{eqnarray}
where the bracket represents the commutator and the anti-commutator is
defined to be $\{A,B\}\equiv AB+BA$; $d_{jkl}$ and $f_{jkl}$ are symmetric
and anti-symmetric structure constants of SU($N$) group, respectively. Any
$N\times N$ Hermitian matrix may be decomposed in terms of these generators
\cite{N-Level-vector}, including quantum states and system observables, i.e.
\begin{eqnarray}
\rho = \frac{1}{N}I + \frac{1}{2} \sum_{j=1}^{N^2-1} p_{j}\lambda_{j}\; , \;
A =\sum_{j=1}^{N^2-1} a_j\lambda_j\; . \label{Eq-rho-A-decompose}
\end{eqnarray}
Here, $p_j = \mathrm{Tr}[\rho \lambda_j]$ and $a_j =\frac{1}{2}
\mathrm{Tr}[A\lambda_j]$. In this form, the $N\times N$ Hermitian matrices
$A$ and $\rho$ may be represented by the $N^2-1$ dimensional real vectors
$\vec{a}$ and $\vec{p}$ with the components of $a_j$ and $p_j$. This is
known as the Bloch vector form of the Hermitian matrix \cite{N-bloch} and
the norms of vectors $\vec{a}$ and $\vec{p}$ are $ |\vec{a}|^2 = \frac{1}{2}\mathrm{Tr}[A^2]$ and $|\vec{p}\,|^2 = 2(\mathrm{Tr}[\rho^2] - 1/N)$, respectively. The quantity $|\vec{p}\,|^2$ may be regarded as a measure of the degree of pureness of quantum state. For pure state $|\vec{p}\,|^2=2(1-1/N)$, while for completely mixed state $|\vec{p}\,|^2 = 0$.

Substituting (\ref{Eq-rho-A-decompose}) into (\ref{Eq-variance-def}), the
variance of observable $A$ for the quantum state $\rho$ may be rewritten as
\begin{eqnarray}
\Delta A^2 & = & \frac{2}{N}|\vec{a}|^2 + \vec{a}\,' \cdot \vec{p}
-|\vec{a}\cdot\vec{p}\,|^2 \; , \label{Eq-Deviation-express}
\end{eqnarray}
with $a_{l}' \equiv \sum_{j,k=1}^{N^2-1} a_ja_kd_{jkl}$. The variance now is
completely characterized by the angles between the vector $\vec{p}$
associated with the quantum state $\rho$ and vectors $\vec{a}\,'$ and
$\vec{a}$ associated with the Hermitian operator $A$, i.e., $\cos\theta_{pa}
\equiv \vec{a}\cdot \vec{p}/(|\vec{a}\,||\vec{p}\,|)$, $\cos\theta_{pa'}
\equiv \vec{a}\,'\cdot \vec{p}/(|\vec{a}\,'||\vec{p}\,|)$. In the Appendix A
section, a general configuration for Bloch vectors of variances will be
given.

\begin{figure}\centering
\scalebox{0.8}{\includegraphics{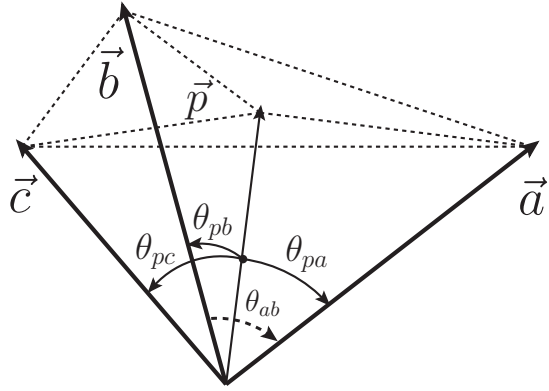}}
\caption{The geometric relation between the quantum
state $\vec{p}$ and the observables $\vec{a}$, $\vec{b}$, and $\vec{c}$ in
3-dimensional real space. The angles between $\vec{p}$ and $\vec{a}$ and
$\vec{b}$ satisfy $|\theta_{pa}-\theta_{pb}| \leq \theta_{ab} \leq
\theta_{pa}+\theta_{pa}$. There are only two free angles in $\theta_{pa}$,
$\theta_{pb}$, and $\theta_{pc}$ because $\vec{a}$, $\vec{b}$, and $\vec{c}$
are 3-dimensional real vectors.}
\label{Fig-2-3-angles}
\end{figure}

In 2-dimensional Hilbert space, the Bloch vector forms of the quantum state
$\rho$ and observables $A$, $B$, and $C$ may be represented by
3-dimensional real vectors $\vec{p}$, $\vec{a}$, $\vec{b}$, $\vec{c}$,
respectively. In this case, the SU(2) generators $\lambda_j$ are Pauli
matrices $\sigma_{i}$, $i=1,2,3$. From equation
(\ref{Eq-Deviation-express}), the variances of $A$, $B$, and $C$ become
\begin{eqnarray}
\Delta A^2 &=& |\vec{a}\,|^2 - |\vec{a}\,|^2|\vec{p}\,|^2\cos^2\theta_{pa} \; ,\label{Eq-delta-a}\\
\Delta B^2 &=& |\vec{b}\,|^2 - |\vec{b}\,|^2|\vec{p}\,|^2\cos^2\theta_{pb} \; ,\label{Eq-delta-b}\\
\Delta C^2 &=& |\vec{c}\,|^2 - |\vec{c}\,|^2|\vec{p}\,|^2\cos^2\theta_{pc} \; .\label{Eq-delta-c}
\end{eqnarray}
Here, $\theta_{pa}$, $\theta_{pb}$, and $\theta_{pc}$ are the angles between
$\vec{a}$, $\vec{b}$, $\vec{c}$, and $\vec{p}$, see Figure
\ref{Fig-2-3-angles}. Note that $\vec{a}\,'=\vec{b}\,'=\vec{c}\,'=0$ due to
the fact that the symmetric structure constants are all zero in SU(2). As
the inversion of a vector, e.g., $\vec{a} \to -\vec{a}$, does not change
the value of the observable variance, we may choose
$\theta_{pa},\theta_{pb}, \theta_{pc} \in [0,\pi/2]$.

\subsection{ The uncertainty relations for general qubit systems}

For two observables $A$ and $B$ and quantum state $\rho$, there exist
the following triangular inequalities for $\theta_{pa}$, $\theta_{pb}$, and
$\theta_{ab}$ (the angle between $\vec{a}$ and $\vec{b}$, see Figure
\ref{Fig-2-3-angles}):
\begin{eqnarray}
|\theta_{pa}-\theta_{pb}| \leq \theta_{ab} \leq
\theta_{pa}+\theta_{pb}\; .
\label{Eq-triangle-inequality}
\end{eqnarray}
Performing cosine to equation (\ref{Eq-triangle-inequality}) and using
equations (\ref{Eq-delta-a}) and (\ref{Eq-delta-b}), we have the following
theorem:
\begin{theorem}
For a qubit system, there exists the following uncertainty relation for
arbitrary observables $A$ and $B$
\begin{eqnarray}
\sqrt{a^2(p^2-1)+ \Delta A^2} \cdot \sqrt{b^2(p^2-1)+\Delta B^2} \geq
\left| \sqrt{a^2-\Delta A^2} \cdot \sqrt{b^2-\Delta B^2} - gp^2 \right| \; ,
\label{Eq-qubit-AB}
\end{eqnarray}
with $a^2 = |\vec{a}\,|^2 = \mathrm{Tr}[A^2]/2$, $b^2= |\vec{b}\,|^2 =
\mathrm{Tr}[B^2]/2$, $p^2=|\vec{p}\,|^2=2(\mathrm{Tr}[\rho^2]-1/2)$, and $g
= \vec{a}\cdot\vec{b} = \mathrm{Tr}[AB]/2$. \label{Theorem-2-vectors}
\end{theorem}
The theorem applies to both pure and mixed states of the qubit system, and
the equality may be obtained when the Bloch vector of the quantum state
$\vec{p}$ is coplane with that of the observables $\vec{a}$ and $\vec{b}$.

\begin{figure}\centering
\scalebox{0.75}{\includegraphics{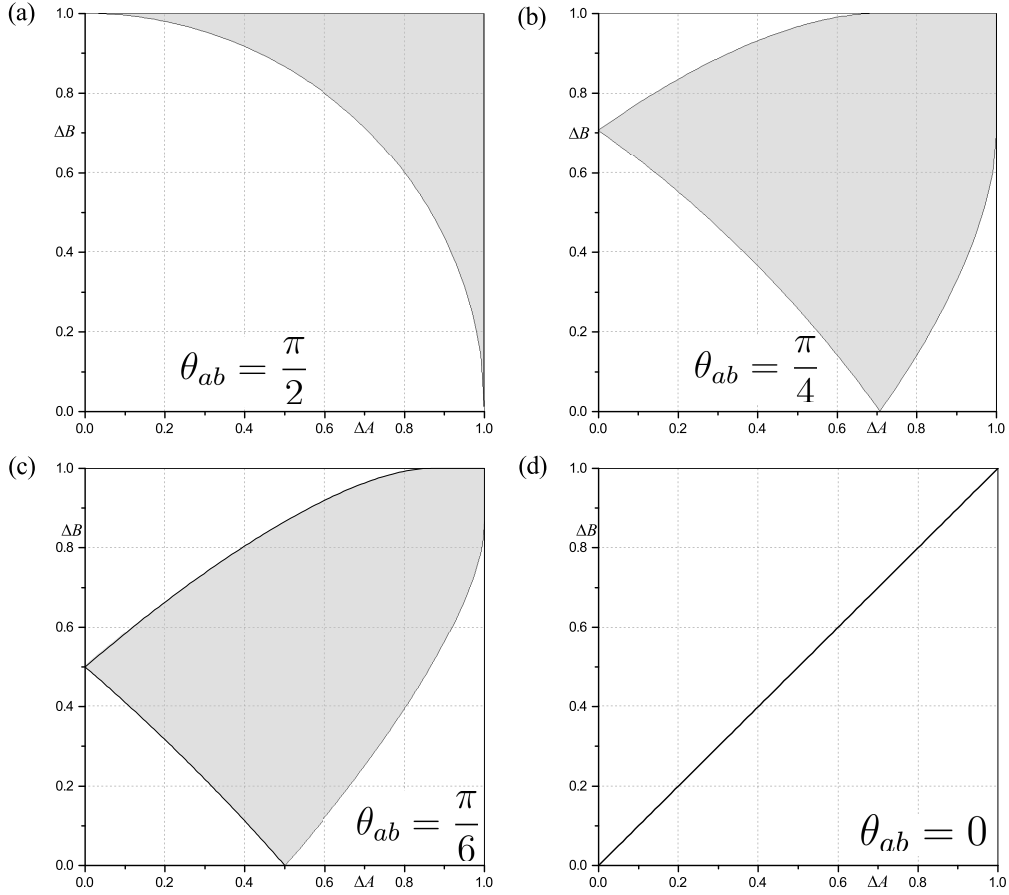}}
\caption{The trade-off relations between the
variances of $A = \vec{\sigma}\cdot \vec{n}_a$ and $B = \vec{\sigma}\cdot
\vec{n}_b$ for different angles $\theta_{ab}$ between $\vec{n}_a$ and
$\vec{n}_b$. The shaded regions correspond to the allowed values of the
variances when (a) $\theta_{ab}=\pi/2$, (b) $\theta_{ab}=\pi/4$, (c)
$\theta_{ab}=\pi/6$, and (d) $\theta_{ab} = 0$ where a line of $\Delta B
=\Delta A$ is obtained.}
\label{Fig-shaded}
\end{figure}

For the completely mixed states of $\rho = I/2$, we have $p^2=0$ and
equation (\ref{Eq-qubit-AB}) leads to
\begin{eqnarray}
\sqrt{\Delta A^2 - a^2} \cdot \sqrt{\Delta B^2 - b^2}
\geq \sqrt{a^2 - \Delta A^2} \cdot \sqrt{b^2 - \Delta B^2} \; .
\end{eqnarray}
This is equivalent to that $\Delta A^2 = a^2$, $\Delta B^2 = b^2$. For pure
states of $p^2=1$, equation (\ref{Eq-qubit-AB}) reduces to
\begin{eqnarray}
\Delta A\Delta B \geq
\left| \sqrt{a^2-\Delta A^2} \cdot \sqrt{b^2-\Delta B^2} - g \right| \; .
\end{eqnarray}
If we further assume $A = \vec{\sigma}\cdot \vec{n}_{a}$ and $B =
\vec{\sigma}\cdot \vec{n}_b$, where $\vec{n}_a$ and $\vec{n}_b$ are
arbitrary unit vectors, then
\begin{eqnarray}
\Delta A\Delta B \geq \left|\sqrt{1-\Delta A^2} \cdot \sqrt{1-\Delta B^2}
-\cos\theta_{ab} \right| \; . \label{Eq-nanb}
\end{eqnarray}
Here $\theta_{ab}$ is the angle between $\vec{n}_a$ and $\vec{n}_b$. Figure
\ref{Fig-shaded} illustrates the trade-off relations between the variances
of $\Delta A^2$ and $\Delta B^2$ for four different values of $\theta_{ab}$.

To compare with the existing uncertainty relations in the market, we exploit
a recent appeared uncertainty relation with state dependent lower bound as
an example \cite{Pati-uncertainty}. It reads,
\begin{eqnarray}
\Delta A^2 + \Delta B^2 \geq \pm i\langle \psi| [A,B]|\psi\rangle +
|\langle \psi|A\pm iB|\psi^{\perp}\rangle|^2 \; , \label{Eq-compare}
\end{eqnarray}
with $\langle \psi|\psi^{\perp}\rangle=0$. Suppose in qubit system
$A=\vec{\sigma} \cdot \vec{n}_a$, $B = \vec{\sigma}\cdot \vec{n}_b$, $\Delta
A^2 = 1/4$ and the angles between observables $A$ and $B$ is $\pi/6$,
equation (\ref{Eq-compare}) then tells
\begin{eqnarray}
\Delta B^2 \geq \pm i\langle \psi| [A,B]|\psi\rangle +
|\langle \psi|A\pm iB|\psi^{\perp}\rangle|^2 -\frac{1}{4} \; .\label{pati-comp}
\end{eqnarray}
While our constraint relation (\ref{Eq-nanb}) gives
\begin{eqnarray}
0 \leq \Delta B \leq \frac{\sqrt{3}}{2} \; , \label{li-comp}
\end{eqnarray}
which can be read directly from Figure \ref{Fig-shaded}(c). Our result
(\ref{li-comp}) is quantum state independent and gives not only the lower
bound, but also a span for $\Delta B$, which is obviously superior to
(\ref{pati-comp}). Moreover, generally speaking the equation
(\ref{Eq-compare}) is not applicable to mixed states $\rho =
\sum_{i=1}^{N}p_i|\psi_i\rangle\langle \psi_i|$ where $p_i>0$ and
$\sum_{i=1}^{N} p_i=1$, since there is no quantum state that could be
orthogonal to all the $|\psi_i\rangle$ in $N$-dimensional system
\cite{Pati-uncertainty}.

When three obervables $A$, $B$, and $C$ are considered, their variances
under the quantum state $\rho$ are characterized by three angles
$\theta_{pa}$, $\theta_{pb}$, and $\theta_{pc}$, see also Figure
\ref{Fig-2-3-angles}. Because only two of these angles are free in
3-dimensional real space, we have the following proposition:
\begin{proposition}
For three independent observables in 2-dimensional Hilbert space, the
trade-off relation for the variances of observables turn out to be an
equality.\label{Proposition-3observables}
\end{proposition}
As the validity of Proposition \ref{Proposition-3observables} is quite
obvious, we only present a simple example as a demonstration of proof.
Suppose three observables are $A=\sigma_1$, $B=\sigma_1\cos\theta_{ab} +
\sigma_2\sin\theta_{ab}$, and $C=\sigma_3$, or in the Bloch vector form
$\vec{a}=(1,0,0)$, $\vec{b} = (\cos\theta_{ab},\sin\theta_{ab},0)$ and
$\vec{c} = (0,0,1)$. An arbitrary quantum state may be constructed as
$\vec{p} = |\vec{p}\, |(\sin\theta\cos\phi, \sin\theta\sin\phi,\cos\theta)$,
where $\theta$, $\phi$ are the polar and azimuthal angles in the
3-dimensional real space. For pure states of $|\vec{p}\, | = 1$,
substituting the values of $\cos\theta_{pa}$, $\cos\theta_{pb}$, and
$\cos\theta_{pc}$ into equations (\ref{Eq-delta-a}), (\ref{Eq-delta-b}) and
(\ref{Eq-delta-c}), one then has the following trade-off relation for $\Delta A^2$, $\Delta B^2$, and $\Delta C^2$,
\begin{eqnarray}
\Delta A^2 + \Delta B^2 +\Delta C^2\sin^2\theta_{ab} +
2\cos\theta_{ab} \sqrt{1-\Delta A^2}\sqrt{1-\Delta B^2} = 2 \; .
\label{Eq-three-correlated}
\end{eqnarray}
Here, $\theta_{ab}$ will always be a constant as long as the observable $B$ is given. It is interesting to observe that the uncertainty relation of equation (\ref{Eq-nanb}) can be obtained by projecting the ``certainty'' relation (\ref{Eq-three-correlated}) onto the $\Delta A$-$\Delta B$ plane with $0\leq \Delta C^2\leq 1$.

In general, by expressing quantum states and physical observables in
Bloch space, the state independent uncertainty relation involving several
observables may be constructed. For the pure qubit system, the variances of
incompatible observables (not only pairwise) cannot be zero simultaneously,
due to the fact that the quantum state of system in the vector form
$\vec{p}$ cannot simultaneously parallel to those unparallel vectors
(incompatible observables, $\vec{a}$, $\vec{b}$, and $\vec{c}$). This
pictorial illustration is quite instructive, which gives the succinct
geometrical account for the uncertainty relations of variances.

\section{Discussion}

There are two types of relations pertaining to the uncertainty principle,
i.e., the uncertainty relation and the measurement disturbance relation
(MDR) \cite{MDR-PRA}. While the uncertainty relation involves the ensemble
properties of variances, the MDR relates the measurement precision to its
back actions, which is currently a hot topic \cite{MDR-PRL}. Though being
fundamentally different, the uncertainty relation and the MDR are both shown
to be correlated with the quantum nonlocality
\cite{ascertaining-correlation, MDR-correlations}. Therefore, it is expected
that every new forms of uncertainty relations may shed some light on the study of the connection between uncertainty principle and quantum
nonlocality.

To summarize, we presented a new type of uncertainty relation which
completely characterizes the trade-off relations among the variances of
several physical obervables for both pure and mixed quantum systems. It
provides the state independent optimal bounds not only for the variances of
pairwise incompatible observables, but also for the multiple incompatible
observables. Unlike the prevailing uncertainty relations in the literature,
our bounds for the variances of observables are immune from the
``triviality'' problem of having null expectation value. As a heuristic
example, we showed, geometrically, that our uncertainty relation turns out
to be an equality for variances of 3 independent observables in
2-dimensional Hilbert space, and pairwise inequalities are merely the
corresponding projections of this equality, which looks enlightening for the
understanding of the complementarity principle in QM.

\vspace{1cm}
\noindent {\Large \bf Acknowledgments}

\noindent This work was supported in part by Ministry of Science and
Technology of the People's Republic of China (2015CB856703), and by the
National Natural Science Foundation of China(NSFC) under the grants
11175249, 11375200, and 11205239.

\newpage

\appendix{\bf \LARGE Appendix}

\section{ General configurations of the Bloch vectors for variances}

\noindent The generators of SU($N$), represented as $\lambda_j$, are $N^2-1$
traceless Hermite matrices satisfying the following relation
\begin{eqnarray}
\lambda_{j} \lambda_k & = & \frac{2}{N}\delta_{jk}I +
\sum_{l=1}^{N^2-1}(if_{jkl}+d_{jkl})\lambda_l \; ,\nonumber
\end{eqnarray}
where $f_{jkl}$ and $d_{jkl}$ are the anti-symmetric and symmetric structure
constants of SU($N$). In term of Bloch vectors, the variance of a physical
observable $A$ takes the following form
\begin{eqnarray}
\Delta A^2 & = & \frac{2}{N}|\vec{a}|^2 +\vec{a}\,' \cdot \vec{p}
-|\vec{a}\cdot\vec{p}|^2 \; . \nonumber
\end{eqnarray}
Here the new vector $\vec{a}\,'$ has the components of $a'_l =
\sum_{j,k=1}^{N^2 -1 } a_ja_kd_{jkl}$ . We may define a new Hermitian
operator $A'\equiv \sum_{l}a'_l\lambda_l$. For given pair of observables $A$
and $B$, there are the vector quaternary \{$\vec{a}$, $\vec{a}\,'$,
$\vec{b}$, $\vec{b}\,'$\}, where
\begin{eqnarray}
|\vec{a}|^2 = \frac{1}{2} \mathrm{Tr}[A^2] \; , \;
|\vec{a}\,'|^2 = \frac{1}{2} \mathrm{Tr}[A'^2]
= \frac{1}{2}(\mathrm{Tr}[A^4] - \frac{1}{N}\mathrm{Tr}[A^2]^2) \; ,\nonumber \\
|\vec{b}|^2 = \frac{1}{2} \mathrm{Tr}[B^2] \; , \;
|\vec{b}\,'|^2 = \frac{1}{2} \mathrm{Tr}[B'^2]
= \frac{1}{2}(\mathrm{Tr}[B^4] - \frac{1}{N}\mathrm{Tr}[B^2]^2) \; . \nonumber
\end{eqnarray}
The angles among the set \{$\vec{a}$, $\vec{a}\,'$, $\vec{b}$,
$\vec{b}\,'$\} are all determined when $A$ and $B$ are given, i.e.,
\begin{eqnarray}
\vec{a} \cdot \vec{b} & = & \frac{1}{2} \mathrm{Tr}[AB] \; , \;
\vec{a} \cdot \vec{b}\,' = \frac{1}{2} \mathrm{Tr}[AB^2] \; , \;
\vec{a} \cdot \vec{a}\,' = \frac{1}{2}\mathrm{Tr}[A^3] \; ,\nonumber \\
\vec{b} \cdot \vec{b}\,' & = & \frac{1}{2} \mathrm{Tr}[B^3] \;\; , \;
\vec{b}\cdot \vec{a}\,' = \frac{1}{2} \mathrm{Tr}[A^2B] \; , \nonumber \\
\vec{b}\,'\cdot \vec{a}\,'  & = &
\frac{1}{2}(\mathrm{Tr}[A^2B^2] - \frac{1}{N}\mathrm{Tr}[A^2]\mathrm{Tr}[B^2]) \; . \nonumber
\end{eqnarray}
Similarly, when there are $k$ observables in $N$-dimensional Hilbert space,
$2k$ vectors in $N^2-1$ real space are obtained with predetermined length
and relative angles.

\section{ An example of Proposition 1}

\noindent Suppose the three observables in 2-dimensional Hilbert space are
$A = \sigma_1$,  $B = \sigma_2$, and $C = \sigma_3$. For quantum state with
$\vec{p} = |\vec{p}\,|(\sin\theta\cos\phi, \sin\theta\sin\phi,\cos\theta)$,
we have
\begin{eqnarray}
\cos\theta_{pa} = \sin\theta\cos\phi \; , \;
\cos\theta_{pb} = \sin\theta\sin\phi \; , \;
\cos\theta_{pc} =\cos\theta \; .\nonumber
\end{eqnarray}
As $\cos^2\theta_{pa} + \cos^2\theta_{pb} + \cos^2\theta_{pc} = 1$, taking
equations (\ref{Eq-delta-a}), (\ref{Eq-delta-b}), and (\ref{Eq-delta-c}) we
have
\begin{eqnarray}
\Delta A^2 +\Delta B^2 + \Delta C^2 = 3-|\vec{p}\,|^2 \; .\nonumber
\end{eqnarray}
The sum of variances of $A$, $B$, $C$ are 2 for pure states and 3 for
completely mixed state.

\section{ An example of $N$-dimensional system}

\noindent For the sake of simplicity and illustration, here we present an
example of state independent trade-off relations for two observables $A$ and
$B$ of $N$-dimension with the Bloch vectors satisfying $\vec{p}\cdot \vec{a}
= \vec{p} \cdot \vec{b}=0$ . This corresponds to the case of $\langle
A\rangle = \langle B\rangle =0$. The variances now become
\begin{eqnarray}
\Delta A^2 & = & \frac{2}{N}|\vec{a}|^2 + |\vec{a}\,'||\vec{p}\,|\cos\theta_{pa'} \; , \\
\Delta B^2 & = & \frac{2}{N}|\vec{b}|^2 + |\vec{b}'||\vec{p}\,|\cos\theta_{pb'} \; .
\end{eqnarray}
Along the same line as equation (\ref{Eq-qubit-AB}), we have the following
trade-off relations between $A$ and $B$ for arbitrary state
\begin{eqnarray}
\sqrt{a'^2|\vec{p}\,|^2- x^2} \cdot \sqrt{b'^2|\vec{p}\,|^2-y^2}
\geq \left| xy-g'|\vec{p}\,|^2 \right|\; . \label{Eq-N-2variables}
\end{eqnarray}
Here, $a'^2 = |\vec{a}\,'|^2$, $b'^2 = |\vec{b}\,'|^2$, $g' = \vec{a}\,'
\cdot \vec{b}\,' $, $x = \Delta A^2 - \frac{1}{N}\mathrm{Tr}[A^2]$, $y =
\Delta B^2 - \frac{1}{N}\mathrm{Tr}[B^2]$. For completely mixed state where
$|\vec{p}\,|=0$, we have $x=y=0$ from equation (\ref{Eq-N-2variables}), and
the variances reduce to $\Delta A^2 = \mathrm{Tr}[A^2]/N$ and $\Delta B^2 =
\mathrm{Tr}[B^2]/N$.


\begin{thebibliography}{99}

\bibitem{Heisenberg-o} W. Heisenberg, \"Uber den anschaulichen Inhalt der
    quantentheoretischen Kinematik und Mechanik. {\it Z. Phys.} {\bf
    43,} 172 (1927); in {\it Quantum theory and Measurement},
    edited by J. A. Wheeler and W. H. Zurek (Princeton University press, Princeton, NJ, 1983), pp.
    62-84.

\bibitem{Robertson} H. P. Robertson, The uncertainty principle. {\it Phys.
    Rev.} {\bf 34,} 163-164 (1929).

\bibitem{Deutsch-entropy} D. Deutsch, Uncertainty in quantum measurements.
    {\it Phys. Rev. Lett.} {\bf 50,} 631-633 (1983).

\bibitem{entropic-un} I. Bia{\l}ynicki-Birula and J. Mycielski, Uncertainty
    relations for information entropy in wave mechanics.
    {\it Commun. Math. Phys.} {\bf 44,} 129-132 (1975).

\bibitem{Maassen-Uffink} H. Maassen and J. B. M. Uffink, Generalized
    entropic uncertainty relations.
    {\it Phys. Rev. Lett.} {\bf 60,} 1103-1106 (1988).


\bibitem{uncertainty-sep} O. G\"{u}hne, Characterizing entanglement via
    uncertainty relations. {\it Phys. Rev. Lett.} {\bf 92,} 117903 (2004).

\bibitem{entropic-nonlocality} J. Oppenheim and S. Wehner, The uncertainty
    principle determines the nonlocality of quantum mechanics. {\it Science} {\bf
    330,} 1072-1074 (2010).

\bibitem{ascertaining-correlation} Jun-Li Li, Kun Du, and Cong-Feng Qiao,
    Ascertaining the uncertainty relations via quantum correlations. {\it J.
    Phys. A} {\bf 47,} 085302 (2014).


\bibitem{Entropy-survey} S. Wehner and A. Winter, Entropic uncertainty
    relations-a survey. {\it New J. Phys.} {\bf 12,} 025009 (2010).


\bibitem{Mixed-improve} Y-M. Park, Improvement of uncertainty relations for
    mixed states. {\it J. Math. Phys.} {\bf 46,} 042109 (2005).

\bibitem{geo-state} H. Heydari and O. Andersson, Geometric uncertainty
    relations for quantum ensembles. {\it Phys. Scr.} {\bf 90}, 025102
    (2015).

\bibitem{Schrodinger-uncertainty} E. Schr\"odinger, About Heisenberg
    uncertainty relation. {\it Sitzungsber. Preuss. Akad. Wiss. Berlin
    (Math. Phys.)} {\bf 19,} 296-303 (1930) (see also, arXiv: quant-ph/9903100).


\bibitem{Pati-uncertainty} L. Maccone and A. K. Pati, Stronger uncertainty
    relations for all incompatible observables. {\it Phys. Rev. Lett.}
    {\bf 113,} 260401 (2014).


\bibitem{Bannur} V. M. Bannur, Comments on ``Stronger Uncertainty
    Relations for All Incompatible Observables''. arXiv: 1502.04853.

\bibitem{geo-ent} G.M. Bosyk, T.M. Os\'{a}n, P.W. Lamberti, and M. Portesi,
    Geometric formulation of the uncertainty principle.
    {\it Phys. Rev. A} {\bf 89,} 034101 (2014).

\bibitem{Variance-entropic} Yichen Huang, Variance-based uncertainty
    relations. {\it Phys. Rev. A} {\bf 86,} 024101 (2012).


\bibitem{N-Level-vector} F. T. Hioe and J. H. Eberly, $N$-Level coherence
    vector and higher conservation laws in quantum optics and quantum
    mechanics.
    {\it Phys. Rev. Lett.} {\bf  47,} 838-841 (1981).


\bibitem{N-bloch} G. Kimura, The Bloch vector for $N$-level systems.
    {\it Phys. Lett. A} {\bf 314,} 339-349 (2003).


\bibitem{MDR-PRA} M. Ozawa, Universally valid reformulation of the
    Heisenberg uncertainty principle on noise and disturbance in measurement.
    {\it Phys. Rev. A} {\bf 67,} 042105 (2003).

\bibitem{MDR-PRL} P. Busch, P. Lahti, and R. F. Werner, Proof of
    Heisenberg's
    error-disturbance relation.
    {\it Phys. Rev. Lett.} {\bf 111,} 160405 (2013).

\bibitem{MDR-correlations} Jun-Li Li, Kun Du, and Cong-Feng Qiao,
    Connection between measurement disturbance relation and multipartite quantum
    correlation.
    {\it Phys. Rev. A} {\bf 91,} 012110 (2015).

\end{thebibliography}
\end{document}